\documentclass[]{interact}

\usepackage{epstopdf}
\usepackage[caption=false]{subfig}

\usepackage[numbers,sort&compress,merge]{natbib}
\bibpunct[, ]{[}{]}{,}{n}{,}{,}
\renewcommand\bibfont{\fontsize{10}{12}\selectfont}

\theoremstyle{plain}

\theoremstyle{definition}

\theoremstyle{remark}

\usepackage{amsmath}
\allowdisplaybreaks

\usepackage{array}

\begin{document}


\title{Fast and Efficient Calculations of Structural Invariants of Chirality}

\author{
\name{He Zhang\textsuperscript{a,b},\thanks{CONTACT He Zhang. Email: zhanghe@ict.ac.cn} Hanlin Mo\textsuperscript{a,b}, You Hao\textsuperscript{a,b}, Shirui Li\textsuperscript{a,b} and Hua Li\textsuperscript{a,b}}
\affil{\textsuperscript{a}Key Laboratory of Intelligent Information Processing, Institute of Computing Technology, Chinese Academy of Sciences, Beijing, China;
	\\ \textsuperscript{b}University of Chinese Academy of Sciences, Beijing, China}
}

\maketitle

\begin{abstract}
Chirality plays an important role in physics, chemistry, biology, and other fields. It describes an essential symmetry in structure. However, chirality invariants are usually complicated in expression or difficult to evaluate. In this paper, we present five general three-dimensional chirality invariants based on the generating functions. And the five chiral invariants have four characteristics:(1) They play an important role in the detection of symmetry, especially in the treatment of ``false zero'' problem. (2) Three of the five chiral invariants decode an universal chirality index. (3) Three of them are proposed for the first time. (4) The five chiral invariants have low order $(\leq4)$, brief expression, low time complexity $(O(n))$ and can act as descriptors of three-dimensional objects in shape analysis. The five chiral invariants give a geometric view to study the chiral invariants. And the experiments show that the five chirality invariants are effective and efficient, they can be used as a tool for symmetry detection or features in shape analysis.
\end{abstract}


\section{Introduction}

Reflection and rotation are two kinds of generally symmetry, which consists in many fields, such as physics, chemistry, biology, art and so on. The reflection symmetry means that the object is divided by a plane or a line into two parts, and one part is the mirror image of another. The rotation symmetry means that the object coincides with itself after rotation.

However, most objects in the world do not have the features above. Chirality is a concept which is used to express the geometric property of an object, it indicates that an object could not be superimposed on its mirror image by translation, scaling and rotation operation. Otherwise, the object is achiral \cite{ruch1977chiral}. The chiral object and its mirror image are called enantiomorph.

It is natural for us to think about how to determine if an object is chiral with an efficient and simple method. Obviously, it is essential to find some metrical expressions that could be used to give a label to the object, for example, achiral or chiral. Furthermore, it is very important for us to discriminate the enantiomorph if the object is not achiral, because the function of the chiral object and its mirror object maybe different even opposite. For example, molecules in chemistry are divided into two types, achiral or chiral, and the handedness of chiral object could be measured with prescriptive resolutions. Actually there are many different methods to measure the chirality in different disciplines \cite{buda1992quantifying,petitjean2003chirality}. An intuitionistic thinking is to compare the two objects and quantify if they are enantiomorph. However, this kind of way ignores the fact that it is usually complex to find the mirror plane which is indispensable in the process of comparison. It makes the problem hard because this kind of way needs us to seek out all possible mirror plane in advance, which is generally time-consuming when the scale of objects is large. This could be understood as that finding out all solutions is usually harder than confirming a solution. It is usually complex and time-consuming to discriminate chirality although many different ideas have been reported, such as searching possible reflective symmetry plane \cite{podolak2006planar}, using general moment \cite{martinet2006accurate} and some other methods \cite{shen1999symmetry,hel1991characterization,loy2006detecting,mitra2006partial,xu2009partial,sun19973d}. And the idea of solving spherical harmonic expression \cite{martinet2006accurate} makes an improvement in three-dimensional situations. 

The concept of geometric invariant cores was proposed in \cite{xu2008geometric}, the construction method could be valid in any degree and any order. Recently, two generating functions, which could re-express the moment invariants and give us a geometric view to consider the inner structure in shape analysis, were shown in \cite{li2017shape}. Furthermore, the study of chiral moment invariant of three-dimensional objects \cite{osipov1995new,hattne2011moment,li2017reflection} gives us another way to judge the chirality of objects. Osipov et al. gave the expression of universal chirality index $G_{0}$ in \cite{osipov1995new} with the integration of four points, and the complexity of $G_{0}$ is $O(n^4)$. By choosing $a=0$, $b=-2$, a chiral invariant (CI) was given in \cite{hattne2011moment} with the complexity is $O(n)$.

In order to simplify the expression $G_{0}$ and then find more essential expressions in particular case, and find out more chiral invariants that are fast and efficient in practice, we decode three chiral invariants from the expression $G_{0}$ and find two other chiral invariants, inspired by the generating functions in \cite{li2017shape} and the propositions in \cite{li2017reflection}. In this paper, we will show five chiral invariants, whose degree and order are no more than 4, with the complexity is $O(n)$. The experiments show that the five chiral invariants are efficient in the discrimination of chirality in three-dimensional situations.

\section{Low order moment chiral invariants}

\subsection{3-D Moments}

Given the density function $\rho(x,y,z)$ of the 3-D object and the order $l+m+n$, the Riemann integral expression defines the 3-D moments as below.
\begin{equation}
M_{lmn}=\int_{ - \infty }^{ + \infty }\int_{ - \infty }^{ + \infty }\int_{ - \infty }^{ + \infty }{x^{l}y^{m}z^{n}\rho(x,y,z)dxdydz}.
\end{equation}

The moments of all order, which are determined by $\rho(x,y,z)$, exist if the density function is bounded and piecewisely continuous in a finite region of 3-D Euclidean space \cite{sadjadi1980three}.

The centroid of the 3-D object could be determined by the zeroth and first-order moments as below.
\begin{equation}
\overline{x}=\frac{M_{100}}{M_{000}},\overline{y}=\frac{M_{010}}{M_{000}},\overline{z}=\frac{M_{001}}{M_{000}}.
\end{equation}

The definition of central moments is
\begin{equation}
\mu_{lmn}=\int_{ - \infty }^{ + \infty }\int_{ - \infty }^{ + \infty }\int_{ - \infty }^{ + \infty }{(x-\overline{x})^{l}(y-\overline{y})^{m}(z-\overline{z})^{n}\rho(x,y,z)dxdydz}.
\end{equation}

The central moments are invariants under the translation operation. Assuming that the centroid of the 3-D object has been moved to the origin of 3-D Euclid space and the object is scaled with $\lambda$, the expressions of central moments of the scaled object and the original object satisfy:
\begin{equation}
\mu_{lmn}^{'}=\int_{ - \infty }^{ + \infty }\int_{ - \infty }^{ + \infty }\int_{ - \infty }^{ + \infty }{x^{l}y^{m}z^{n}\rho(\frac{x}{\lambda},\frac{y}{\lambda},\frac{z}{\lambda})dxdydz}=\lambda^{l+m+n+3}\mu_{lmn}.
\end{equation}

Furthermore, dividing the central moments by the $\mu_{000}$ with designated order when the calculations of the central moments finish, the result we get would be invariable under uniform scaling for 3-D objects \cite{lo19893}, it is
\begin{equation}
\eta_{lmn}=\frac{\mu_{lmn}}{\mu_{000}^{1+(l+m+n)/3}}.
\end{equation}

Now we know that the expression (5) is an invariant under translation and uniform scaling. It maybe natural for us to think about what is the form of moment invariant under the rotation which is an important part in similarity transformation, but we will skip this step and take the form of chiral invariant into consideration directly, since we could choose the mirror plane in any direction.

\subsection{3-D Invariants and Generating functions}

Four invariant geometric primitives for invariants under translation and rotation in 3-D Euclidean space were proposed in \cite{xu2008geometric}, they are the distance $D(i,j)$, the area $A(i,j,k)$, the dot product $D_{p}(i,j,k)$ and the signed volume $V(i,j,k,l)$. The dot-product function $f(i,j)$ and the cross-product function $g(i,j,k)$ were shown in \cite{li2017shape} as the generating functions in 3-D Euclidean space. The expressions of them are 
\begin{equation}
f(i,j)=(x_{i},y_{i},z_{i})\cdot(x_{j},y_{j},z_{j})=x_{i}x_{j}+y_{i}y_{j}+z_{i}z_{j},
\end{equation}
\begin{equation}
\begin{aligned}
g(i,j,k)=\left|\begin{array}{cccc}
x_{i}&y_{i}&z_{i}\\
x_{j}&y_{j}&z_{j}\\
x_{k}&y_{k}&z_{k}
\end{array}\right|=&x_{i}y_{j}z_{k}+x_{j}y_{k}z_{i}+x_{k}y_{i}z_{j}\\&-x_{i}y_{k}z_{j}-x_{j}y_{i}z_{k}-x_{k}y_{j}z_{i},
\end{aligned}
\end{equation}
where (6) is the dot-product of two vectors and (7) is the determinant of matrix which constructed by three vectors. By combining different (6) and (7) and choosing multiple integrals carefully, we could get the moment invariants with their expression are the multiple integrals of the multiplication of generating functions. And the composite expressions of (6) and (7) is called the primitive invariants (PIs). 

For example, the expressions of invariants proposed in \cite{sadjadi1980three} are as follows.
\begin{equation}
\begin{aligned}
J_{1}&=\mu_{200}+\mu_{020}+\mu_{002}\\
J_{2}&=\mu_{200}\mu_{020}\mu_{002}+2\mu_{110}\mu_{101}\mu_{011}-\mu_{011}^{2}\mu_{200}-\mu_{110}^{2}\mu_{002}--\mu_{101}^{2}\mu_{020}\\
J_{3}&=\mu_{020}\mu_{002}-\mu_{011}^{2}+\mu_{200}\mu_{002}-\mu_{101}^{2}+\mu_{200}\mu_{020}-\mu_{110}^{2}
\end{aligned}
\end{equation}
The relationship between the expressions and the generating functions of 3-D Euclidean space are shown as bellows.
\begin{equation}
\begin{aligned}
J_{1}&\Leftrightarrow f(1,1)\\
J_{2}&\Leftrightarrow g(1,2,3)^2\\
J_{3}&\Leftrightarrow f(1,1)f(2,2)-f(1,2)^2 
\end{aligned}
\end{equation}
Assuming that the centroid of the 3-D object has been moved to the origin of 3-D Euclidean space and taking $J_{1}$ for instance, the first expression in (9) means that 
\begin{equation}
\begin{aligned}
&\int_{ - \infty }^{ + \infty }\int_{ - \infty }^{ + \infty }\int_{ - \infty }^{ + \infty }f(1,1)\rho(x,y,z)dxdydz\\
=&\int_{ - \infty }^{ + \infty }\int_{ - \infty }^{ + \infty }\int_{ - \infty }^{ + \infty }(x_{1}x_{1}+y_{1}y_{1}+z_{1}z_{1})\rho(x,y,z)dxdydz\\
=&\int_{ - \infty }^{ + \infty }\int_{ - \infty }^{ + \infty }\int_{ - \infty }^{ + \infty }x_{1}^{2}\rho(x,y,z)dxdydz+\\
&\int_{ - \infty }^{ + \infty }\int_{ - \infty }^{ + \infty }\int_{ - \infty }^{ + \infty }y_{1}^{2}\rho(x,y,z)dxdydz+\\
&\int_{ - \infty }^{ + \infty }\int_{ - \infty }^{ + \infty }\int_{ - \infty }^{ + \infty }z_{1}^{2}\rho(x,y,z)dxdydz\\
=&\mu_{200}+\mu_{020}+\mu_{002}\\
=&J_{1}.
\end{aligned}
\end{equation}

\subsection{Chiral Invariants}

The expression of universal chirality index $G_{0}$, which is the integration of four points, was given by Osipov et al. in \cite{osipov1995new}, and the complexity of $G_{0}$ is $O(n^4)$. The expression of $G_{0}$ is as bellows.
\begin{equation}
G_{0}=\int\frac{(\mathbf r_{12}\times \mathbf r_{34}\cdot \mathbf r_{14})(\mathbf r_{12}\cdot \mathbf r_{23})(\mathbf r_{23}\cdot \mathbf r_{34})}{(r_{12}r_{23}r_{34})^{a}r_{14}^{b}}\rho(r_{1})\rho(r_{2})\rho(r_{3})\rho(r_{4})dr_{1}dr_{2}dr_{3}dr_{4}
\end{equation}
And $\mathbf r_{1}$, $\mathbf r_{2}$, $\mathbf r_{3}$ and $\mathbf r_{4}$ are four points in 3-D Euclid space, $\mathbf r_{ij}=\mathbf r_{i}-\mathbf r_{j}$, $r_{ij}=\Arrowvert \mathbf r_{ij}\Arrowvert$, $a$ and $b$ are arbitrary integers. Actually there are many different choices of $a$ and $b$, and different choices lead to different results. For example, $G_{0}$ would be a scale invariant by choosing $a=2$ and $b=1$ \cite{osipov1995new}, and the expression would be zero if chose $a=0$ and $b=0$. Hattne and Lamzin showed a chiral invariant in \cite{hattne2011moment} by choosing $a=0$, $b=-2$ in $G_{0}$, and the complexity of the chiral invariant is $O(n)$. The choice of $a$ and $b$ in \cite{hattne2011moment} could be considered as a balance between computational efficiency and robustness.

With choosing $a=0$ and $b=-2$, we expand the expression 
\begin{equation}
\frac{(\mathbf r_{12}\times \mathbf r_{34}\cdot \mathbf r_{14})(\mathbf r_{12}\cdot \mathbf r_{23})(\mathbf r_{23}\cdot \mathbf r_{34})}{(r_{12}r_{23}r_{34})^{a}r_{14}^{b}}.
\end{equation} 
The result is a combination of 192 monomials, each of which is composed by 3 $f(i,j)$s and 1 $g(i,j,k)$. We convert the 192 monomials into the expressions that composed by $\mu_{lmn}$, just like the process in (10), the result shows that some of them are equal to zero and some of them are equal or opposite to other monomials. Moreover, some of them contain the $\mu_{lmn}$ which is zero in the context of central moments. We remove the monomials with the above characteristics from the  192 monomials, and get three chiral invariants. The expressions of them are listed as below.

\begin{equation}
S_{1}=f(1,1)f(1,2)f(2,3)g(1,2,3)
\end{equation}
\begin{equation}
S_{2}=f(1,1)f(1,2)f(3,3)g(1,2,3)
\end{equation}
\begin{equation}
S_{3}=f(1,2)f(1,3)f(2,4)g(1,3,4)
\end{equation}
After adjusting the order of the points, we find (13) is opposite to the first chiral invariant proposed in \cite{li2017reflection}, and (14) is equal to the second chiral invariant in \cite{li2017reflection}.

The analysis about the structure of the chiral invariants was proposed in \cite{li2017reflection}, it gives the guiding principle about how to construct a new chiral invariant. Moreover, the comparison in \cite{xu2008geometric} shown that the moment invariants of lower orders or lower degrees are more stable than the moment invariants of higher orders or higher degrees, and the former is usually more time-saving than the later as a result of the multinomials of the later are more complicated and the size are bigger. Therefore, we find two another chiral invariants with the order and the degree of them are no more than four. The expressions of them are listed as below.
\begin{equation}
S_{4}=f(1,1)f(2,3)^{2}g(1,2,3)
\end{equation}
\begin{equation}
S_{5}=f(1,2)g(1,2,3)g(1,3,4)^{2}
\end{equation}

The fully expanded expressions of $S_{1}$, $S_{2}$, $S_{3}$, $S_{4}$, $S_{5}$ are given in appendix A.

\subsection{Analysis of the five chiral invariants}

\subsubsection{Structure of the five chiral invariants}

The total number of points that participate in the integral is called as the degree of the invariant, and the highest occurrence number of the points is called the order of the invariant. Apart from the degree and the order, the number of generating functions $f(i,j)$ and $g(i,j,k)$ that compose the invariant are the important property of the moment invariant. A necessary and sufficient condition for a chiral invariant was given in \cite{li2017reflection}, it is obviously that (13) (14) (15) (16) (17) are five chiral invariants with their degree and order are no more than four. The values of relative parameters mentioned above are listed in Table 1.

\begin{table}
	\tbl{The degree, order and the number of $f(i,j)$ and $g(i,j,k)$ that compose the five chiral invariants.}
	{\begin{tabular}{lcccc}  
			\toprule
			Chiral invariant expression & Degree & Order & Number of $f(i,j)$ & Number of $g(i,j,k)$ \\  
			\midrule
			$S_{1}=f(1,1)f(1,2)f(2,3)g(1,2,3)$ & 3 & 4 & 3 & 1 \\
			$S_{2}=f(1,1)f(1,2)f(3,3)g(1,2,3)$ & 3 & 4 & 3 & 1 \\
			$S_{3}=f(1,2)f(1,3)f(2,4)g(1,3,4)$ & 3 & 4 & 3 & 1 \\ 
			$S_{4}=f(1,1)f(2,3)^{2}g(1,2,3)$   & 4 & 3 & 3 & 1 \\ 
			$S_{5}=f(1,2)g(1,2,3)g(1,3,4)^{2}$ & 4 & 4 & 1 & 3 \\ 
			\bottomrule
	\end{tabular}}
	\label{table}
\end{table}

\subsubsection{Functional Independent}

We could use the five chiral invariants to describe the shape of 3-D objects when they are functional independent of each other, which is considered as a more rigid requirement than linear independent. A technique about how to determine the functional independent of a group of functions was proposed by Brown et al. in \cite{brown1935functional}. It could be described as below.

Assuming that there are a group of functions $f_{j}(x_{1},x_{2},\ldots,x_{n})$, and $j=1,\ldots,m$, $x_{1},x_{2},\ldots,x_{n}$ are the variables. And we could deduce a $m\times n$ matrix $\mathbf J$ which is the Jacobian matrix of this group of functions. So this group of functions are functional independent if and only if the rank of the Jacobian matrix $\mathbf J$ is $m$.

Based on the technique mentioned above, we verified that the set of (13) (14) (15) (16) (17) is independent with the help of Maple software.

\subsubsection{Computation Complexity}

The computation complexity of expression $G_{0}$ given by Osipov et al. \cite{osipov1995new} is $O(n^4)$. By choosing $a=0$, $b=-2$, the complexity of the CI \cite{hattne2011moment} is $O(n)$. The complexity of the five chiral invariants shown in (13) (14) (15) (16) (17) is $O(n)$, too. The comparison of CI and the five chiral invariants is shown in Table 2.
 
\begin{table}
	\tbl{The comparison of CI and the five chiral invariants.}
	{\begin{tabular}{ccc}  
			\toprule
			Chiral Invariant & Number of additions & Number of multiplications \\  
			\midrule
			  CI    & 117 & About 110  \\
			$S_{1}$ & 125 & 58 \\
			$S_{2}$ & 143 & 51  \\ 
			$S_{3}$ & 64 & 60  \\ 
			$S_{4}$ & 95 & 108 \\ 
			$S_{5}$ & 151 & 383 \\ 
			\bottomrule
	\end{tabular}}
    \tabnote{Remark: We get the result with the help of simplification command of the Maple.}
	\label{table}
\end{table}

\subsubsection{``False Zero" Problem and Sign}

The ``false zero'' problem is a typical problem\cite{fowler2005quantification}, it means that the value of chiral invariants would be 0 even if the object is chiral and Fowler gave an example in \cite{fowler2005quantification}. The five chiral invariants can not solve this problem. However, \cite{li2017reflection} gave a technique to detect the potential planes of symmetry of an object and then to judge if the two parts divided by the plane are mirrored. The technique is effective and the five chiral invariants are helpful in the verification part. 

In the practice, the sign of the five chiral invariants could be modified with multiplying by -1, if the user has chosen a calibrated object. 

We can use the calibrated model to define the correct sign to normalize them. As well known that the concept of ``left" or ``right" is only relative, there is no absolute, clear standard of it. Some ``physical" or ``empirical" methods are definitely needed.

\section{Experimental Results}

\subsection{Biphenyl}

\begin{figure}
	\centering
	\includegraphics[scale=0.6]{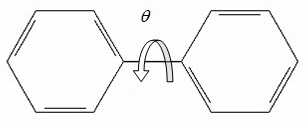}
	\caption{The structure of the biphenyl molecule. The arrow indicates the rotation direction of the right benzene ring alone the C-C bond which links the left benzene and the right benzene.}
	\label{figure}
\end{figure}
Biphenyl is a typical achiral molecule (Figure 1). When the right benzene ring is rotated alone the C-C bond which links the left benzene and the right benzene, the chirality of the structure is determined by the angle of rotation. We get the structure data from the PubChem database of NIH \cite{pcd7095}, and calculate the values of the five chiral invariants and CI (Figure 2). The result shows that the values of the five chiral invariants and CI are zero at $\theta=0^{\circ}$, $\theta=90^{\circ}$ and $\theta=180^{\circ}$. The curves of $S_{1}$, $S_{2}$, $S_{3}$, $S_{4}$ perform sinusoidal (differ by at most a negative sign) like the curve of CI with getting their highest absolute values at $\theta=45^{\circ}$, $\theta=135^{\circ}$. The curve of $S_{5}$ is a little different to others and it gets highest absolute values at $\theta=60^{\circ}$ and $\theta=120^{\circ}$. 
\begin{figure}
	\centering
	\includegraphics[scale=0.35]{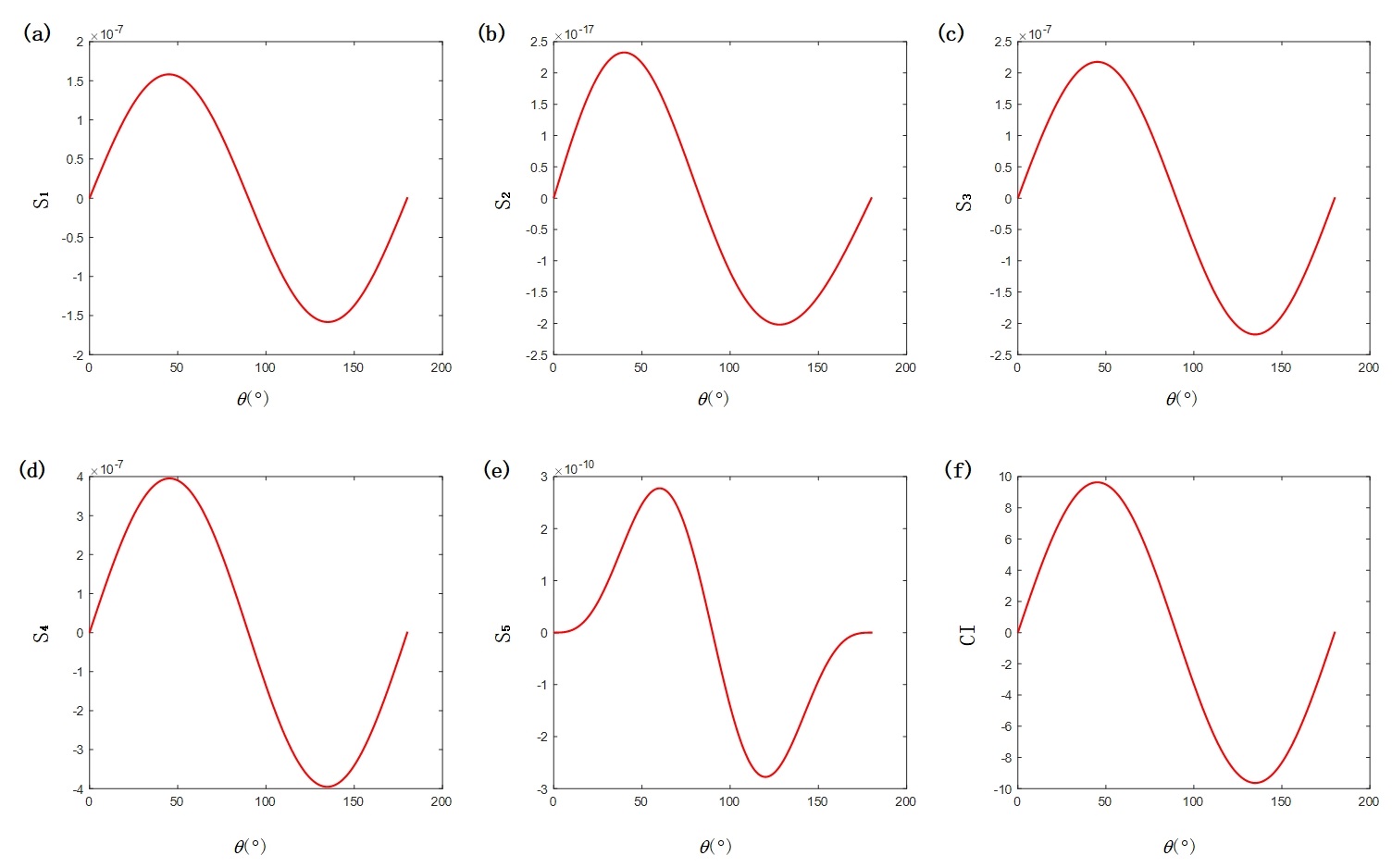}
	\caption{The curve of the values of the five chiral invariants and CI at different angle of rotation on Biphenyl. And (a) is the curve of $S_{1}$, (b) is the curve of $S_{2}$, (c) is the curve of $S_{3}$, (d) is the curve of $S_{4}$, (e) is the curve of $S_{5}$, (f) is the curve of CI. The signs of $S_{1}$ and $S_{3}$ are modified with -1 for a better comparison with CI.}
	\label{figure}
\end{figure}

When adding different degrees of normal noise to the structure data of biphenyl, the experiments show that $S_{1}$, $S_{3}$, $S_{4}$ are robust to normal noise scaled with $10^{-1}$ like CI, $S_{5}$ is robust to normal noise scaled with $10^{-2}$ and $S_{2}$ is robust to normal noise scaled with $10^{-6}$. The curves of the five chiral invariants and CI with adding normal noise to the structure data are shown in Figure 3.
\begin{figure}
	\centering
	\includegraphics[scale=0.35]{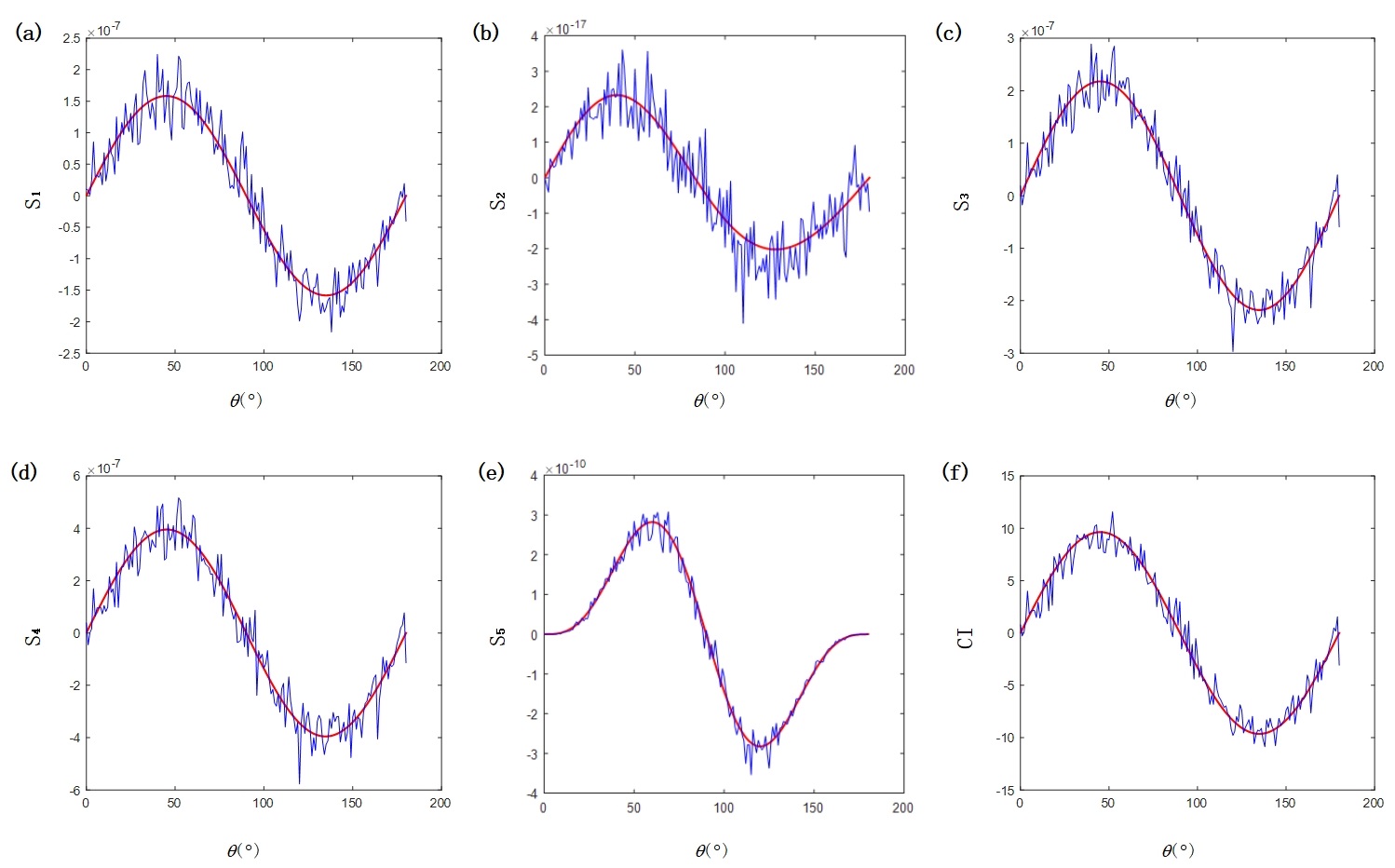}
	\caption{The red curves are the values of the five chiral invariants and CI at different angle of rotation on Biphenyl, and the blue curves are the values of them when adding different degrees of normal noise to the structure data of biphenyl. And the blue curves in (a) (c) (d) (f) are the values of $S_{1}$, $S_{3}$, $S_{4}$, CI with adding normal noise scaled with $10^{-1}$ to the structure data, the blue curve in (e) is the value of $S_{5}$ with adding normal noise scaled with $10^{-2}$ to the structure data, the blue curve in (b) is the value of $S_{2}$ with adding normal noise scaled with $10^{-6}$ to the structure data.}
	\label{figure}
\end{figure}

\subsection{Platonic Objects}

We choose the Platonic Objects, which are obviously achiral, to verify if the five chiral invariants are valid in the symmetry detection. We use the Wolfram Mathematica 11 to get the vertex-coordinates of the Tetrahedron, Cube, Octahedron, Dodecahedron and Icosahedron and then calculate the values of the five chiral invariants and CI. The result is shown in Table 3. 

\begin{table}
	\tbl{The values of the five chiral invariants and CI on Platonic objects.}
	{\begin{tabular}{ >{\centering\arraybackslash} m{2cm}  >{\centering\arraybackslash} m{2cm}  >{\centering\arraybackslash} m{2cm}  >{\centering\arraybackslash} m{2cm} >{\centering\arraybackslash} m{2cm} >{\centering\arraybackslash} m{2cm}}  
			\toprule
			Platonico bjects & Tetrahedron & Cube & Octahedron & Dodecahedron & Icosahedron \\  
			\midrule
			Figure 
			& \includegraphics[scale=0.14]{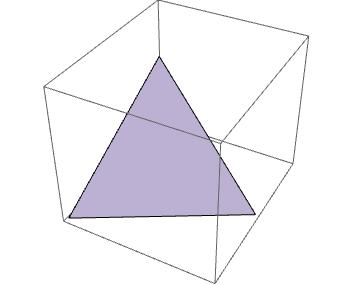} 
			& \includegraphics[scale=0.1]{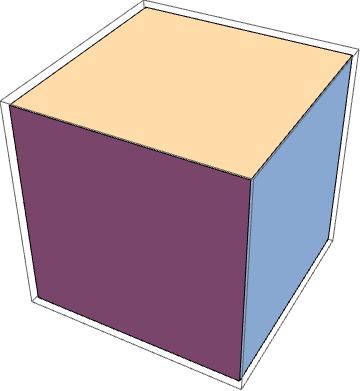}  
			& \includegraphics[scale=0.1]{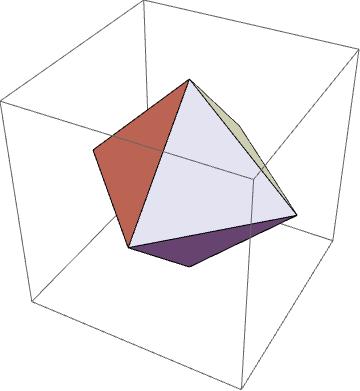} 
			& \includegraphics[scale=0.1]{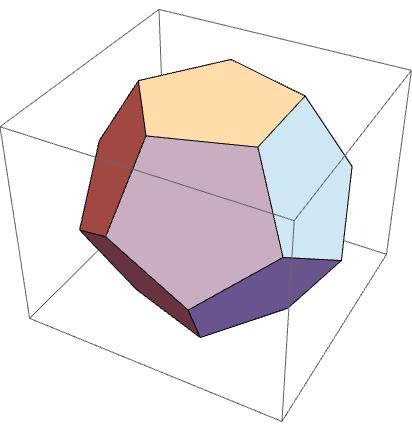}  
			& \includegraphics[scale=0.1]{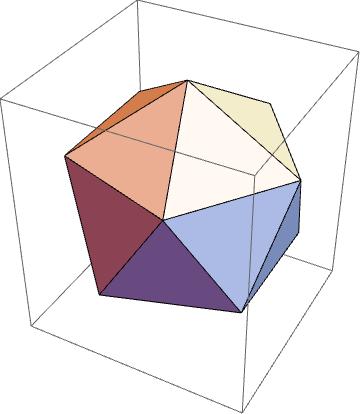} \\  
			\midrule			
			$S_{1}$ & 0 & 0 & 0 & 0 & 0   \\
			$S_{2}$ & 0 & 0 & 0 & 0 & 0   \\ 
			$S_{3}$ & 0 & 0 & 0 & 0 & 0   \\ 
			$S_{4}$ & 0 & 0 & 0 & 0 & 0   \\ 
			$S_{5}$ & 0 & 0 & 0 & 0 & 0   \\ 
			CI    & 0 & 0 & 0 & 0 & 0   \\
			\bottomrule
	\end{tabular}}
	\label{table}
\end{table}
\begin{table}
	\tbl{The error-values of the five chiral invariants and CI on the horse model.}
	{\begin{tabular}{cccccc}  
			\toprule
			Operation & Translation\textsuperscript{a} & Rotation\textsuperscript{b} & Mirror\textsuperscript{c} & Scale(1.5)\textsuperscript{d} & Scale(2)\textsuperscript{e} \\  
			\midrule			
			$S_{1}$ & $5.10\times10^{-12}$ & $2.31\times10^{-12}$ & 0 & 1.36 & 2.87   \\
			$S_{2}$ & $7.47\times10^{-12}$ & $1.90\times10^{-12}$ & 0 & 3.52 & 5.09   \\ 
			$S_{3}$ & $2.83\times10^{-12}$ & $4.03\times10^{-12}$ & 0 & 0.27 & 0.72   \\ 
			$S_{4}$ & $4.74\times10^{-12}$ & $2.12\times10^{-12}$ & 0 & 9.86 & 8.10   \\ 
			$S_{5}$ & $1.67\times10^{-11}$ & $3.14\times10^{-12}$ & 0 & 4.13 & 6.26   \\ 
			  CI    & $9.09\times10^{-12}$ & $3.38\times10^{-12}$ & 0 & 2.79 & 3.53   \\
			\bottomrule
	\end{tabular}}
	\tabnote{\textsuperscript{a}  This means that the model is not scaled, and the operation is only translation.\\
		\textsuperscript{b} This means that the model is not scaled, and the operation is only rotation.\\
		\textsuperscript{c} This means that the model is not scaled, and the operation is only mirror.\\
		\textsuperscript{d} This means that the model is scaled with 1.5, and the operations are translation, rotation and mirror.\\
		\textsuperscript{e} This means that the model is scaled with 2, and the operations are translation, rotation and mirror.}
	\label{table}
\end{table}
\begin{figure}
	\centering
	\includegraphics[scale=0.4]{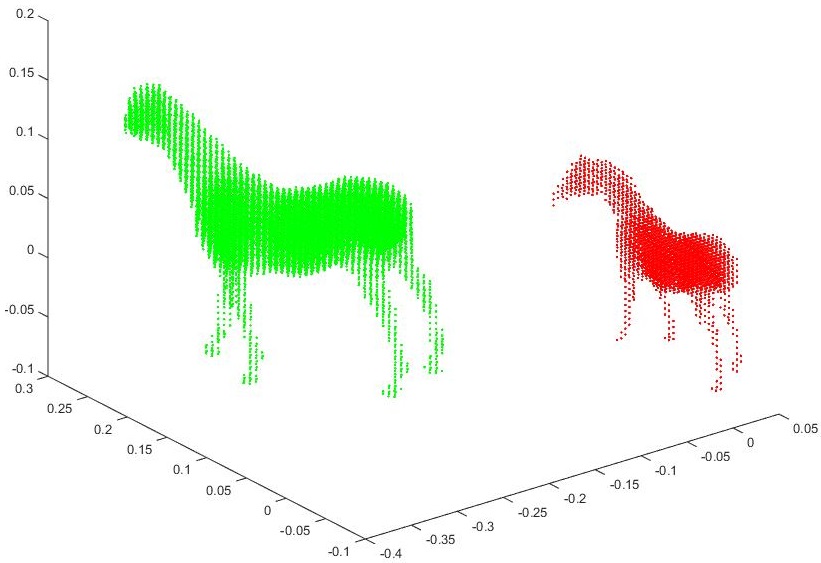}
	\caption{The result after operations mentioned above on the horse model. The red horse is the original voxel model, the green horse is the voxel model experiences the translation, rotation and mirror operation after the horse model is scaled with $1.5$. The figure is just like the big green horse is turning her head to right to look back to the little red horse, the red horse is turning her head to the left to look forward to the big green horse.}
	\label{figure}
\end{figure}

\subsection{Horse Model}

The horse model is a typical chiral object. We use the method in \cite{karabassi1999fast} to get the voxel data on the horse model in different scales with the step=$0.05$.

The experiments show that the absolute values of the five chiral invariants almost do not change in translation, rotation and mirror operation and change slightly in scale operation. And the sign of the values only change in mirror operation.

The way to evaluate relative error is
\begin{equation}
e_{i}=\frac{\vert S_{i}\vert - \vert S_{i}^{'}\vert}{\vert S_{i}\vert + \vert S_{i}^{'}\vert}\times 100\%.
\end{equation}
Where $e_{i}$ is the relative error, $S_{i}$ is the value of the chiral invariants, $S_{i}^{'}$ is the value of the same chiral invariants after relative operation, $\vert\ \ \vert$ means getting the absolute value.

For example, we set translation vector $=(0.1, 0.3, 0.05)$, rotation vector $= (0, 0, 135)$, mirror vector $= (0, 1, 0)$. These vectors mean that the scaled model is translated with $(0.1, 0.3, 0.05)$, rotated $135$ degrees around the $z$ axis and mirrored with the normal vector of the mirror plane is $(0, 1, 0)$. The values of the relative error on the five chiral invariants are shown in Table 4. The result after the operations on horse model which is scaled with $1.5$ is shown in Figure 4.
\subsection{``False Zero" Object}

In order to show the availability of the five chiral invariants in the task of symmetry detection, we conducted the following experiments.

Firstly, we construct a simple 3-D object with the values of the five chiral invariants are 0. We fix the four points (-1,0,0), (1,-2,0), (1,2,0), (-1,2,0), and they located on the bottom of the 3-D object. The fifth point move from (-20,1,1) to (20,1,1) with the step length is 0.05, the density of these five points is (1,1,2,1,1). The process is shown in figure 5. The values of the five chiral invariants in above process are shown in figure 6, the result shows that the values of the five chiral invariants and CI experience the process from positive to negative, and they are 0 when the offset in about [222,571]. We choose offset=405 as an example, and the position of the fifth point is (0.3,1,1). The 3-D object is shown in figure 7-(a).

Secondly, we use the technique proposed in \cite{li2017reflection} to show that the five chiral invariants are helpful in the verification part. We set k=4 in $M^{k}(\varphi,\theta)$, and then get potential symmetry planes of the object, and the normals of them are shown in table 5. The figure 7-(b) shows the object and a potential symmetry plane with the No.7 normal in table 5. The plane divide the 3-D object into two parts, the upper object and the lower object. Then we calculate the values of the five chiral invariants, and the result is shown in table 6. The result shows that the two parts are not enantiomorph. After verifying all potential symmetry planes, we get the conclusion that the zeros are false. Moreover, we get the same conclusion at other offset values.

This is an example about the application of the five chiral invariants in the task of symmetry detection.

\begin{figure}
	\centering
	\includegraphics[scale=0.2]{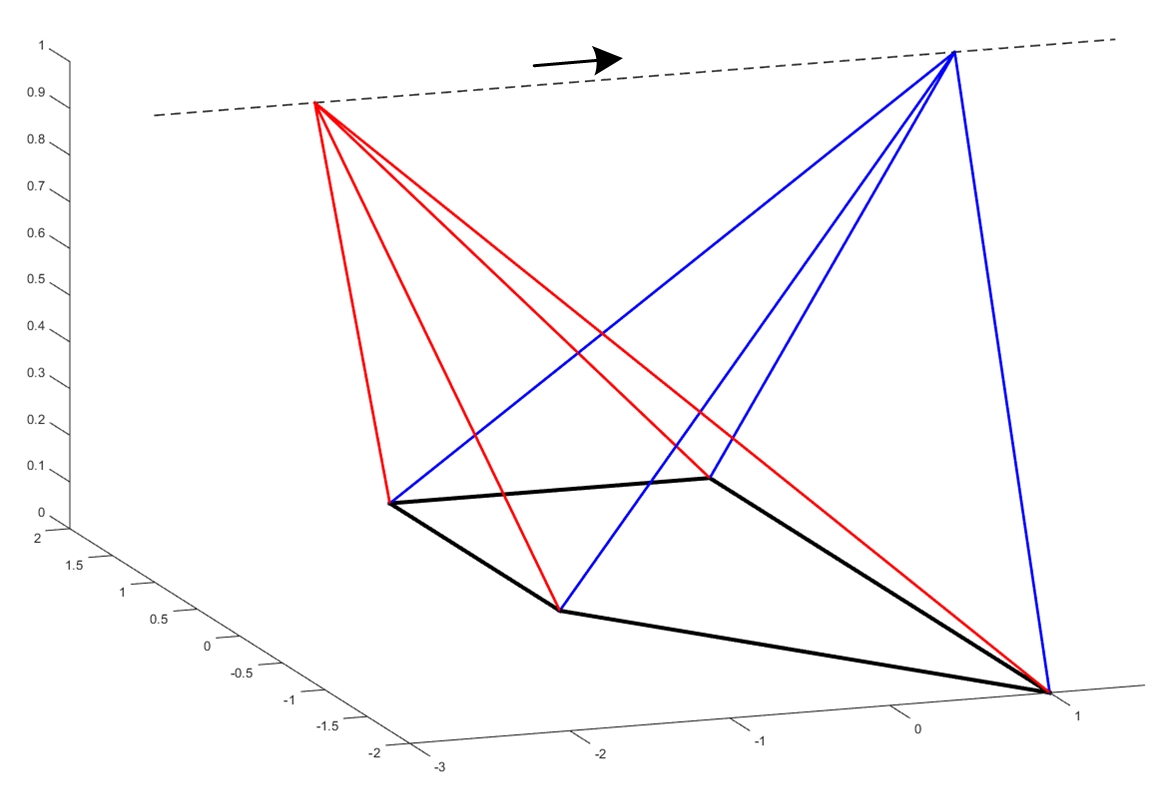}
	\caption{The construction process of chiral object with the values of the five chiral invariants are 0.}
	\label{figure}
\end{figure}

\begin{figure}
	\centering
	\includegraphics[scale=0.3]{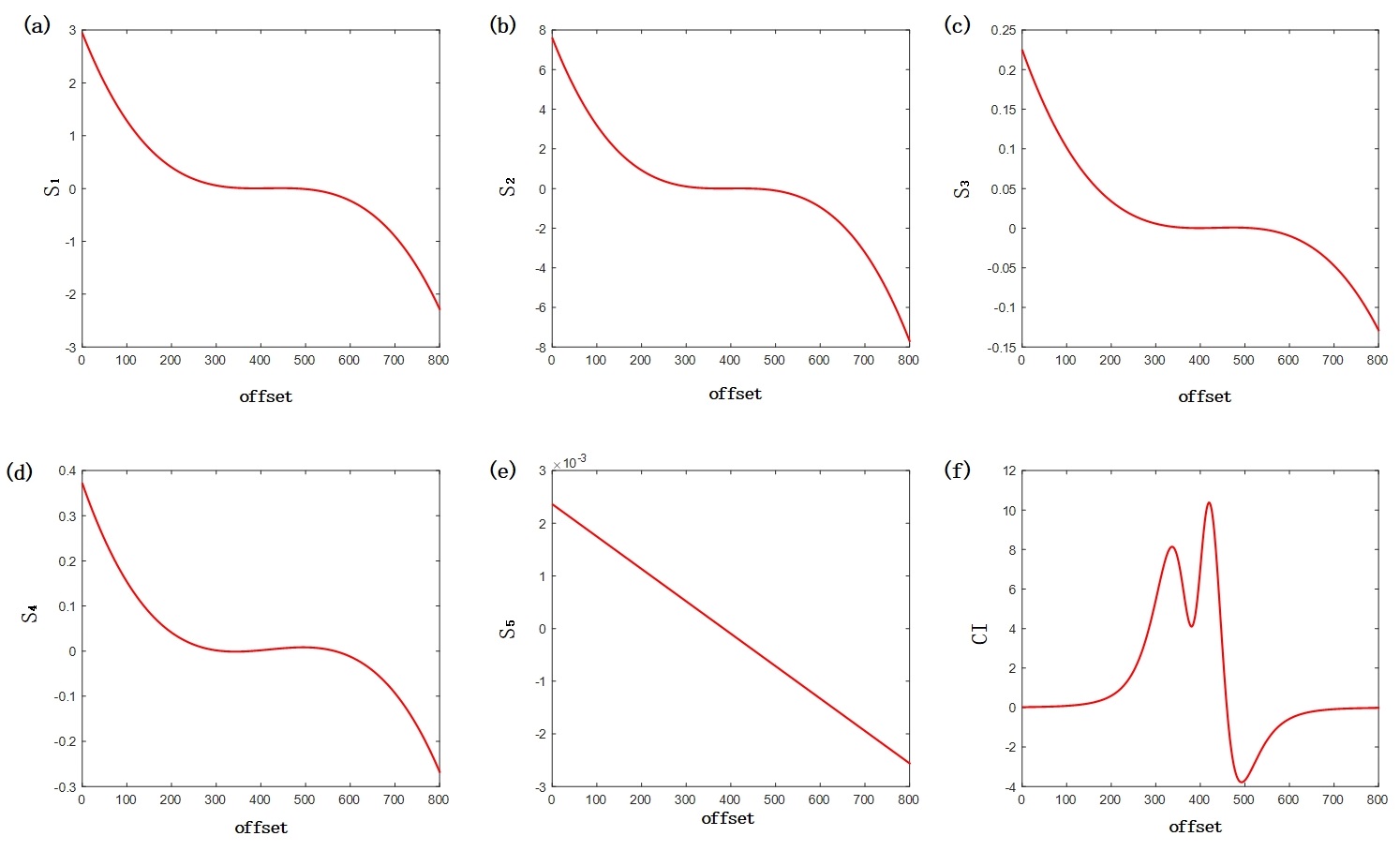}
	\caption{The values of the five chiral invariants in the construction process. The signs of $S_{1}$, $S_{3}$ and $S_{4}$ are motified with -1 in order to clearly show the changing process.}
	\label{figure}
\end{figure}

\begin{figure}
	\centering
	\includegraphics[scale=0.2]{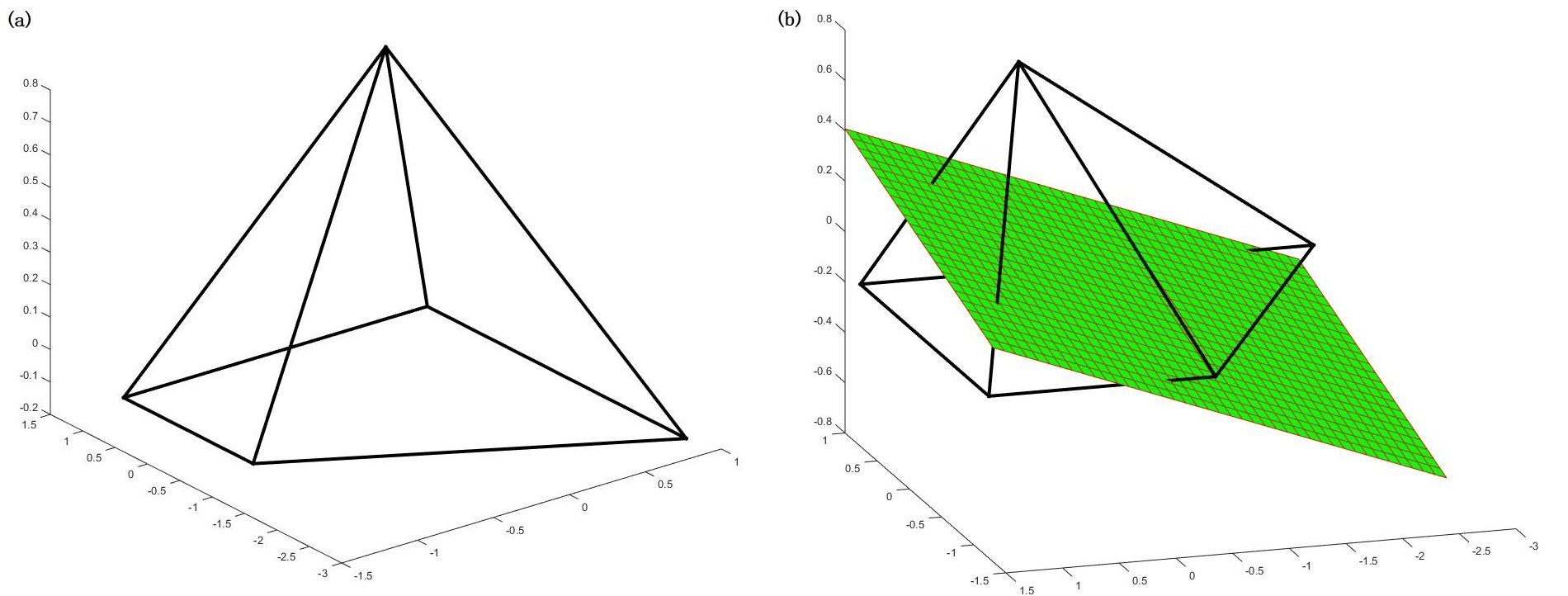}
	\caption{The left figure (a) is the ``False Zero" chiral object, the right figure (b) is the ``False Zero" chiral object and a potential symmetry.}
	\label{figure}
\end{figure}

\begin{table}
	\tbl{The normals of potential symmetry planes.}
	{\begin{tabular}{cccccccccccc}  
			\toprule
			NO. & 1 & 2 & 3 & 4 & 5 & 6 & 7 & 8 & 9 & 10 &$\cdots$ \\  
			\midrule			
			$x$ & 0 & 0.9511 & 0.9489 & 0.8074 &  0.1788 & -0.3777 &-0.1517 &-0.2440 &0.1517  & 0.3777 & \\
			$y$ & 0 & -0.0502 & 0.0801 & 0.5691 & -0.1435 & -0.3266 &-0.1639  &0.9685 &0.1639 &0.3266& $\cdots$ \\ 
			$z$ & 1 & 0.3049 & 0.3051 & 0.1553 & 0.9734 & 0.8664 &0.9747 &0.0505 &-0.9747 &-0.8664 &   \\ 
			\bottomrule
	\end{tabular}}
	\label{table}
\end{table}

\begin{table}
	\tbl{The values of the five chiral invariants on the upper object and lower object in figure 7-(b).}
	{\begin{tabular}{cccccc}  
			\toprule
		    $S$	& $S_{1}$ & $S_{2}$ & $S_{3}$ & $S_{4}$  & $S_{5}$  \\  
			\midrule			
			Upper object& $-3.3789\times10^{-5}$ & $3.2856\times10^{-5}$ & $2.8561\times10^{-5}$ & $-5.1065\times10^{-5}$ & $1.0687\times10^{-8}$  \\
			Lower object& $-0.0017$ & $0.0023$ & $-6.6493\times10^{-5}$ & $-0.0012$ & $3.0568\times10^{-5}$     \\ 
			\bottomrule
	\end{tabular}}
	\label{table}
\end{table}

\section{Conclusion}
We have shown that the universal chirality index $G_{0}$ in specified circumstance could be decoded into more essential expressions $S_{1}$, $S_{2}$ and $S_{3}$. As the expressions proposed for the first time, $S_{3}$, $S_{4}$ and $S_{5}$ perform as well as $S_{1}$ and $S_{2}$. The five chiral invariants have brief expression with low order $(\leq4)$ and low time complexity $(O(n))$.

With regard to a three-dimensional chiral object $A$ and its mirror counterpart $A^{'}$, the signs of the same chirality invariant will be opposite and the absolute values of the same chirality invariant will be equal. And we have shown they play an important
role in the detection of symmetry, especially in the treatment of ``false zero" problem.

The five chirality invariants are effective and efficient in experiments. They give a geometric view to study the chiral invariants and could be used as a group of descriptors in the task of shape analysis.

\section*{Acknowledgment}

This work was partly funded by National Natural Science Foundation of China (Grant No.60573154, 61227802 and 61379082).

\bibliographystyle{tfo}
\bibliography{interacttfosample}

\appendix
\section{The expressions of the five chiral invariants}

\begin{equation*}
\begin{aligned}
S_{1}=&\ \ \  \eta_{002}\eta_{012}\eta_{103}+\eta_{002}\eta_{012}\eta_{121}+\eta_{002}\eta_{012}\eta_{301}-\eta_{002}\eta_{013}\eta_{102}+\eta_{002}\eta_{021}\eta_{112}
\\&+\eta_{002}\eta_{021}\eta_{130}+\eta_{002}\eta_{021}\eta_{310}-\eta_{002}\eta_{022}\eta_{111}-\eta_{002}\eta_{031}\eta_{102}-\eta_{002}\eta_{040}\eta_{111}
\\&-\eta_{002}\eta_{102}\eta_{211}+\eta_{002}\eta_{111}\eta_{202}+\eta_{002}\eta_{111}\eta_{400}-\eta_{002}\eta_{112}\eta_{201}-\eta_{002}\eta_{130}\eta_{201}
\\&-\eta_{002}\eta_{201}\eta_{310}-\eta_{003}\eta_{011}\eta_{103}-\eta_{003}\eta_{011}\eta_{121}-\eta_{003}\eta_{011}\eta_{301}+\eta_{003}\eta_{013}\eta_{101}
\\&+\eta_{003}\eta_{031}\eta_{101}+\eta_{003}\eta_{101}\eta_{211}+\eta_{004}\eta_{011}\eta_{102}-\eta_{004}\eta_{012}\eta_{101}+\eta_{004}\eta_{020}\eta_{111}
\\&-\eta_{004}\eta_{021}\eta_{110}+\eta_{004}\eta_{110}\eta_{201}-\eta_{004}\eta_{111}\eta_{200}-\eta_{011}\eta_{012}\eta_{112}-\eta_{011}\eta_{012}\eta_{130}
\\&-\eta_{011}\eta_{012}\eta_{310}+\eta_{011}\eta_{021}\eta_{103}+\eta_{011}\eta_{021}\eta_{121}+\eta_{011}\eta_{021}\eta_{301}+\eta_{011}\eta_{022}\eta_{102}
\\&-\eta_{011}\eta_{022}\eta_{120}+\eta_{011}\eta_{030}\eta_{112}+\eta_{011}\eta_{030}\eta_{130}+\eta_{011}\eta_{030}\eta_{310}-\eta_{011}\eta_{040}\eta_{120}
\\&-\eta_{011}\eta_{102}\eta_{220}-\eta_{011}\eta_{102}\eta_{400}+\eta_{011}\eta_{103}\eta_{201}-\eta_{011}\eta_{112}\eta_{210}+\eta_{011}\eta_{120}\eta_{202}
\\&+\eta_{011}\eta_{120}\eta_{400}+\eta_{011}\eta_{121}\eta_{201}-\eta_{011}\eta_{130}\eta_{210}+\eta_{011}\eta_{201}\eta_{301}-\eta_{011}\eta_{210}\eta_{310}
\\&+\eta_{012}\eta_{013}\eta_{110}-\eta_{012}\eta_{020}\eta_{103}-\eta_{012}\eta_{020}\eta_{121}-\eta_{012}\eta_{020}\eta_{301}+\eta_{012}\eta_{031}\eta_{110}
\\&+\eta_{012}\eta_{040}\eta_{101}-\eta_{012}\eta_{101}\eta_{202}+\eta_{012}\eta_{101}\eta_{220}+\eta_{012}\eta_{110}\eta_{211}+\eta_{013}\eta_{020}\eta_{120}
\\&-\eta_{013}\eta_{021}\eta_{101}-\eta_{013}\eta_{030}\eta_{110}-\eta_{013}\eta_{101}\eta_{201}+\eta_{013}\eta_{102}\eta_{200}+\eta_{013}\eta_{110}\eta_{210}
\\&-\eta_{013}\eta_{120}\eta_{200}-\eta_{020}\eta_{021}\eta_{112}-\eta_{020}\eta_{021}\eta_{130}-\eta_{020}\eta_{021}\eta_{310}+\eta_{020}\eta_{022}\eta_{111}
\\&+\eta_{020}\eta_{031}\eta_{120}+\eta_{020}\eta_{103}\eta_{210}-\eta_{020}\eta_{111}\eta_{220}-\eta_{020}\eta_{111}\eta_{400}+\eta_{020}\eta_{120}\eta_{211}
\\&+\eta_{020}\eta_{121}\eta_{210}+\eta_{020}\eta_{210}\eta_{301}-\eta_{021}\eta_{031}\eta_{101}+\eta_{021}\eta_{040}\eta_{110}-\eta_{021}\eta_{101}\eta_{211}
\\&-\eta_{021}\eta_{110}\eta_{202}+\eta_{021}\eta_{110}\eta_{220}-\eta_{022}\eta_{101}\eta_{210}+\eta_{022}\eta_{110}\eta_{201}-\eta_{030}\eta_{031}\eta_{110}
\\&-\eta_{030}\eta_{110}\eta_{211}-\eta_{031}\eta_{101}\eta_{201}+\eta_{031}\eta_{102}\eta_{200}+\eta_{031}\eta_{110}\eta_{210}-\eta_{031}\eta_{120}\eta_{200}
\\&-\eta_{040}\eta_{101}\eta_{210}+\eta_{040}\eta_{111}\eta_{200}+\eta_{101}\eta_{102}\eta_{112}+\eta_{101}\eta_{102}\eta_{130}+\eta_{101}\eta_{102}\eta_{310}
\\&+\eta_{101}\eta_{112}\eta_{120}-\eta_{101}\eta_{112}\eta_{300}+\eta_{101}\eta_{120}\eta_{130}+\eta_{101}\eta_{120}\eta_{310}-\eta_{101}\eta_{130}\eta_{300}
\\&-\eta_{101}\eta_{201}\eta_{211}+\eta_{101}\eta_{202}\eta_{210}+\eta_{101}\eta_{210}\eta_{400}-\eta_{101}\eta_{300}\eta_{310}-\eta_{102}\eta_{103}\eta_{110}
\\&-\eta_{102}\eta_{110}\eta_{121}-\eta_{102}\eta_{110}\eta_{301}+\eta_{102}\eta_{200}\eta_{211}-\eta_{103}\eta_{110}\eta_{120}+\eta_{103}\eta_{110}\eta_{300}
\\&-\eta_{103}\eta_{200}\eta_{210}-\eta_{110}\eta_{120}\eta_{121}-\eta_{110}\eta_{120}\eta_{301}+\eta_{110}\eta_{121}\eta_{300}-\eta_{110}\eta_{201}\eta_{220}
\\&-\eta_{110}\eta_{201}\eta_{400}+\eta_{110}\eta_{210}\eta_{211}+\eta_{110}\eta_{300}\eta_{301}-\eta_{111}\eta_{200}\eta_{202}+\eta_{111}\eta_{200}\eta_{220}
\\&+\eta_{112}\eta_{200}\eta_{201}-\eta_{120}\eta_{200}\eta_{211}-\eta_{121}\eta_{200}\eta_{210}+\eta_{130}\eta_{200}\eta_{201}+\eta_{200}\eta_{201}\eta_{310}
\\&-\eta_{200}\eta_{210}\eta_{301};
\\
\\S_{2}=\ &-\eta_{002}\eta_{012}\eta_{103}-\eta_{002}\eta_{012}\eta_{121}-\eta_{002}\eta_{012}\eta_{301}+\eta_{002}\eta_{013}\eta_{102}+\eta_{002}\eta_{013}\eta_{120}
\\&+\eta_{002}\eta_{013}\eta_{300}-\eta_{002}\eta_{030}\eta_{103}-\eta_{002}\eta_{030}\eta_{121}-\eta_{002}\eta_{030}\eta_{301}+\eta_{002}\eta_{031}\eta_{102}
\\&+\eta_{002}\eta_{031}\eta_{120}+\eta_{002}\eta_{031}\eta_{300}+\eta_{002}\eta_{102}\eta_{211}-\eta_{002}\eta_{103}\eta_{210}+\eta_{002}\eta_{120}\eta_{211}
\\&-\eta_{002}\eta_{121}\eta_{210}-\eta_{002}\eta_{210}\eta_{301}+\eta_{002}\eta_{211}\eta_{300}+\eta_{003}\eta_{011}\eta_{103}+\eta_{003}\eta_{011}\eta_{121}
\\&+\eta_{003}\eta_{011}\eta_{301}-\eta_{003}\eta_{013}\eta_{101}+\eta_{003}\eta_{020}\eta_{112}+\eta_{003}\eta_{020}\eta_{130}+\eta_{003}\eta_{020}\eta_{310}
\\&-\eta_{003}\eta_{022}\eta_{110}-\eta_{003}\eta_{031}\eta_{101}-\eta_{003}\eta_{040}\eta_{110}-\eta_{003}\eta_{101}\eta_{211}+\eta_{003}\eta_{110}\eta_{202}
\\&+\eta_{003}\eta_{110}\eta_{400}-\eta_{003}\eta_{112}\eta_{200}-\eta_{003}\eta_{130}\eta_{200}-\eta_{003}\eta_{200}\eta_{310}-\eta_{004}\eta_{011}\eta_{102}
\\&-\eta_{004}\eta_{011}\eta_{120}-\eta_{004}\eta_{011}\eta_{300}+\eta_{004}\eta_{012}\eta_{101}+\eta_{004}\eta_{030}\eta_{101}+\eta_{004}\eta_{101}\eta_{210}
\\&-\eta_{011}\eta_{012}\eta_{112}-\eta_{011}\eta_{012}\eta_{130}-\eta_{011}\eta_{012}\eta_{310}+\eta_{011}\eta_{021}\eta_{103}+\eta_{011}\eta_{021}\eta_{121}
\\&+\eta_{011}\eta_{021}\eta_{301}-\eta_{011}\eta_{030}\eta_{112}-\eta_{011}\eta_{030}\eta_{130}-\eta_{011}\eta_{030}\eta_{310}+\eta_{011}\eta_{040}\eta_{102}
\\&+\eta_{011}\eta_{040}\eta_{120}+\eta_{011}\eta_{040}\eta_{300}-\eta_{011}\eta_{102}\eta_{202}+\eta_{011}\eta_{102}\eta_{220}+\eta_{011}\eta_{103}\eta_{201}
\end{aligned}
\end{equation*}
\begin{equation*}
\begin{aligned}
\\&-\eta_{011}\eta_{112}\eta_{210}-\eta_{011}\eta_{120}\eta_{202}+\eta_{011}\eta_{120}\eta_{220}+\eta_{011}\eta_{121}\eta_{201}-\eta_{011}\eta_{130}\eta_{210}
\\&+\eta_{011}\eta_{201}\eta_{301}-\eta_{011}\eta_{202}\eta_{300}-\eta_{011}\eta_{210}\eta_{310}+\eta_{011}\eta_{220}\eta_{300}+\eta_{012}\eta_{013}\eta_{110}
\\&+\eta_{012}\eta_{022}\eta_{101}+\eta_{012}\eta_{031}\eta_{110}-\eta_{012}\eta_{101}\eta_{220}-\eta_{012}\eta_{101}\eta_{400}+\eta_{012}\eta_{103}\eta_{200}
\\&+\eta_{012}\eta_{110}\eta_{211}+\eta_{012}\eta_{121}\eta_{200}+\eta_{012}\eta_{200}\eta_{301}-\eta_{013}\eta_{020}\eta_{102}-\eta_{013}\eta_{020}\eta_{120}
\\&-\eta_{013}\eta_{020}\eta_{300}-\eta_{013}\eta_{021}\eta_{101}+\eta_{013}\eta_{030}\eta_{110}-\eta_{013}\eta_{101}\eta_{201}+\eta_{013}\eta_{110}\eta_{210}
\\&+\eta_{020}\eta_{021}\eta_{112}+\eta_{020}\eta_{021}\eta_{130}+\eta_{020}\eta_{021}\eta_{310}-\eta_{020}\eta_{031}\eta_{102}-\eta_{020}\eta_{031}\eta_{120}
\\&-\eta_{020}\eta_{031}\eta_{300}-\eta_{020}\eta_{102}\eta_{211}+\eta_{020}\eta_{112}\eta_{201}-\eta_{020}\eta_{120}\eta_{211}+\eta_{020}\eta_{130}\eta_{201}
\\&+\eta_{020}\eta_{201}\eta_{310}-\eta_{020}\eta_{211}\eta_{300}-\eta_{021}\eta_{022}\eta_{110}-\eta_{021}\eta_{031}\eta_{101}-\eta_{021}\eta_{040}\eta_{110}
\\&-\eta_{021}\eta_{101}\eta_{211}+\eta_{021}\eta_{110}\eta_{202}+\eta_{021}\eta_{110}\eta_{400}-\eta_{021}\eta_{112}\eta_{200}-\eta_{021}\eta_{130}\eta_{200}
\\&-\eta_{021}\eta_{200}\eta_{310}+\eta_{022}\eta_{030}\eta_{101}+\eta_{022}\eta_{101}\eta_{210}-\eta_{022}\eta_{110}\eta_{201}+\eta_{030}\eta_{031}\eta_{110}
\\&-\eta_{030}\eta_{101}\eta_{220}-\eta_{030}\eta_{101}\eta_{400}+\eta_{030}\eta_{103}\eta_{200}+\eta_{030}\eta_{110}\eta_{211}+\eta_{030}\eta_{121}\eta_{200}
\\&+\eta_{030}\eta_{200}\eta_{301}-\eta_{031}\eta_{101}\eta_{201}+\eta_{031}\eta_{110}\eta_{210}-\eta_{040}\eta_{110}\eta_{201}+\eta_{101}\eta_{102}\eta_{112}
\\&+\eta_{101}\eta_{102}\eta_{130}+\eta_{101}\eta_{102}\eta_{310}+\eta_{101}\eta_{112}\eta_{120}+\eta_{101}\eta_{112}\eta_{300}+\eta_{101}\eta_{120}\eta_{130}
\\&+\eta_{101}\eta_{120}\eta_{310}+\eta_{101}\eta_{130}\eta_{300}-\eta_{101}\eta_{201}\eta_{211}-\eta_{101}\eta_{210}\eta_{220}-\eta_{101}\eta_{210}\eta_{400}
\\&+\eta_{101}\eta_{300}\eta_{310}-\eta_{102}\eta_{103}\eta_{110}-\eta_{102}\eta_{110}\eta_{121}-\eta_{102}\eta_{110}\eta_{301}-\eta_{103}\eta_{110}\eta_{120}
\\&-\eta_{103}\eta_{110}\eta_{300}+\eta_{103}\eta_{200}\eta_{210}-\eta_{110}\eta_{120}\eta_{121}-\eta_{110}\eta_{120}\eta_{301}-\eta_{110}\eta_{121}\eta_{300}
\\&+\eta_{110}\eta_{201}\eta_{202}+\eta_{110}\eta_{201}\eta_{400}+\eta_{110}\eta_{210}\eta_{211}-\eta_{110}\eta_{300}\eta_{301}-\eta_{112}\eta_{200}\eta_{201}
\\&+\eta_{121}\eta_{200}\eta_{210}-\eta_{130}\eta_{200}\eta_{201}-\eta_{200}\eta_{201}\eta_{310}+\eta_{200}\eta_{210}\eta_{301};
\\
\\S_{3}=&\ \ \ 
\eta_{002}^{2}\eta_{020}\eta_{111}-\eta_{002}^{2}\eta_{021}\eta_{110}+\eta_{002}^{2}\eta_{110}\eta_{201}-\eta_{002}^{2}\eta_{111}\eta_{200}-\eta_{002}\eta_{011}^2\eta_{111}
\\&+2\eta_{002}\eta_{011}\eta_{012}\eta_{110}-\eta_{002}\eta_{011}\eta_{020}\eta_{102}+\eta_{002}\eta_{011}\eta_{020}\eta_{120}+\eta_{002}\eta_{011}\eta_{021}\eta_{101}
\\&-\eta_{002}\eta_{011}\eta_{030}\eta_{110}-\eta_{002}\eta_{011}\eta_{101}\eta_{201}+\eta_{002}\eta_{011}\eta_{102}\eta_{200}+\eta_{002}\eta_{011}\eta_{110}\eta_{210}
\\&-\eta_{002}\eta_{011}\eta_{120}\eta_{200}-\eta_{002}\eta_{012}\eta_{020}\eta_{101}+\eta_{002}\eta_{012}\eta_{101}\eta_{200}-\eta_{002}\eta_{020}^2\eta_{111}
\\&+\eta_{002}\eta_{020}\eta_{021}\eta_{110}+\eta_{002}\eta_{020}\eta_{101}\eta_{210}-\eta_{002}\eta_{020}\eta_{110}\eta_{201}+\eta_{002}\eta_{021}\eta_{110}\eta_{200}
\\&+\eta_{002}\eta_{101}^2\eta_{111}-2\eta_{002}\eta_{101}\eta_{102}\eta_{110}-\eta_{002}\eta_{101}\eta_{110}\eta_{120}+\eta_{002}\eta_{101}\eta_{110}\eta_{300}
\\&-\eta_{002}\eta_{101}\eta_{200}\eta_{210}-\eta_{002}\eta_{110}\eta_{200}\eta_{201}+\eta_{002}\eta_{111}\eta_{200}^2-\eta_{003}\eta_{011}^2\eta_{110}
\\&+\eta_{003}\eta_{011}\eta_{020}\eta_{101}-\eta_{003}\eta_{011}\eta_{101}\eta_{200}+\eta_{003}\eta_{101}^2\eta_{110}+\eta_{011}^3\eta_{102}-\eta_{011}^3\eta_{120}
\\&-\eta_{011}^2\eta_{012}\eta_{101}+\eta_{011}^2\eta_{020}\eta_{111}+\eta_{011}^2\eta_{021}\eta_{110}+\eta_{011}^2\eta_{030}\eta_{101}-2\eta_{011}^2\eta_{101}\eta_{210}
\\&+2\eta_{011}^2\eta_{110}\eta_{201}-\eta_{011}\eta_{012}\eta_{020}\eta_{110}-\eta_{011}\eta_{012}\eta_{110}\eta_{200}-2\eta_{011}\eta_{020}\eta_{021}\eta_{101}
\\&-\eta_{011}\eta_{020}\eta_{101}\eta_{201}+\eta_{011}\eta_{020}\eta_{102}\eta_{200}+\eta_{011}\eta_{020}\eta_{110}\eta_{210}-\eta_{011}\eta_{020}\eta_{120}\eta_{200}
\\&+\eta_{011}\eta_{021}\eta_{101}\eta_{200}+\eta_{011}\eta_{030}\eta_{110}\eta_{200}+\eta_{011}\eta_{101}^2\eta_{102}+2\eta_{011}\eta_{101}^2\eta_{120}
\\&-\eta_{011}\eta_{101}^2\eta_{300}+2\eta_{011}\eta_{101}\eta_{200}\eta_{201}-2\eta_{011}\eta_{102}\eta_{110}^2-\eta_{011}\eta_{102}\eta_{200}^2
\\&-\eta_{011}\eta_{110}^2\eta_{120}+\eta_{011}\eta_{110}^2\eta_{300}-2\eta_{011}\eta_{110}\eta_{200}\eta_{210}+\eta_{011}\eta_{120}\eta_{200}^2
\\&+\eta_{012}\eta_{020}^2\eta_{101}-\eta_{012}\eta_{020}\eta_{101}\eta_{200}-\eta_{012}\eta_{101}^3+2\eta_{012}\eta_{101}\eta_{110}^2-\eta_{020}^2\eta_{101}\eta_{210}
\\&+\eta_{020}^2\eta_{111}\eta_{200}-\eta_{020}\eta_{021}\eta_{110}\eta_{200}+\eta_{020}\eta_{101}\eta_{102}\eta_{110}+2\eta_{020}\eta_{101}\eta_{110}\eta_{120}
\\&-\eta_{020}\eta_{101}\eta_{110}\eta_{300}+\eta_{020}\eta_{101}\eta_{200}\eta_{210}-\eta_{020}\eta_{110}^2\eta_{111}+\eta_{020}\eta_{110}\eta_{200}\eta_{201}
\\&-\eta_{020}\eta_{111}\eta_{200}^2-2\eta_{021}\eta_{101}^2\eta_{110}+\eta_{021}\eta_{110}^3-\eta_{030}\eta_{101}\eta_{110}^2+\eta_{101}^3\eta_{210}
\\&-\eta_{101}^2\eta_{110}\eta_{201}-\eta_{101}^2\eta_{111}\eta_{200}+\eta_{101}\eta_{102}\eta_{110}\eta_{200}+\eta_{101}\eta_{110}^2\eta_{210}
\end{aligned}
\end{equation*}
\begin{equation*}
\begin{aligned}
\\&-\eta_{101}\eta_{110}\eta_{120}\eta_{200}-\eta_{110}^3\eta_{201}+\eta_{110}^2\eta_{111}\eta_{200};
\\
\\S_{4}=&\ \ \ \  \eta_{002}\eta_{012}\eta_{103}-\eta_{002}\eta_{013}\eta_{102}+2\eta_{002}\eta_{021}\eta_{112}-2\eta_{002}\eta_{022}\eta_{111}+\eta_{002}\eta_{030}\eta_{121}
\\&-\eta_{002}\eta_{031}\eta_{120}+2\eta_{002}\eta_{111}\eta_{202}-2\eta_{002}\eta_{112}\eta_{201}+2\eta_{002}\eta_{120}\eta_{211}-2\eta_{002}\eta_{121}\eta_{210}
\\&+\eta_{002}\eta_{210}\eta_{301}-\eta_{002}\eta_{211}\eta_{300}-\eta_{003}\eta_{011}\eta_{103}+\eta_{003}\eta_{013}\eta_{101}-\eta_{003}\eta_{020}\eta_{112}
\\&+\eta_{003}\eta_{022}\eta_{110}-\eta_{003}\eta_{110}\eta_{202}+\eta_{003}\eta_{112}\eta_{200}+\eta_{004}\eta_{011}\eta_{102}-\eta_{004}\eta_{012}\eta_{101}
\\&-\eta_{011}\eta_{012}\eta_{112}+2\eta_{011}\eta_{013}\eta_{111}+\eta_{011}\eta_{021}\eta_{121}-\eta_{011}\eta_{022}\eta_{102}+\eta_{011}\eta_{022}\eta_{120}
\\&+\eta_{011}\eta_{030}\eta_{130}-2\eta_{011}\eta_{031}\eta_{111}-\eta_{011}\eta_{040}\eta_{120}-2\eta_{011}\eta_{102}\eta_{202}+2\eta_{011}\eta_{103}\eta_{201}
\\&+2\eta_{011}\eta_{112}\eta_{210}+2\eta_{011}\eta_{120}\eta_{220}-2\eta_{011}\eta_{121}\eta_{201}-2\eta_{011}\eta_{130}\eta_{210}-\eta_{011}\eta_{201}\eta_{301}
\\&+\eta_{011}\eta_{202}\eta_{300}+\eta_{011}\eta_{210}\eta_{310}-\eta_{011}\eta_{220}\eta_{300}-\eta_{012}\eta_{013}\eta_{110}-2\eta_{012}\eta_{020}\eta_{121}
\\&+2\eta_{012}\eta_{022}\eta_{101}+2\eta_{012}\eta_{031}\eta_{110}+\eta_{012}\eta_{101}\eta_{202}-\eta_{012}\eta_{103}\eta_{200}-2\eta_{012}\eta_{110}\eta_{211}
\\&+2\eta_{012}\eta_{121}\eta_{200}+\eta_{013}\eta_{020}\eta_{102}-2\eta_{013}\eta_{021}\eta_{101}-\eta_{020}\eta_{021}\eta_{130}+2\eta_{020}\eta_{022}\eta_{111}
\\&+\eta_{020}\eta_{031}\eta_{120}-2\eta_{020}\eta_{102}\eta_{211}-2\eta_{020}\eta_{111}\eta_{220}+2\eta_{020}\eta_{112}\eta_{201}+2\eta_{020}\eta_{121}\eta_{210}
\\&-\eta_{020}\eta_{201}\eta_{310}+\eta_{020}\eta_{211}\eta_{300}-2\eta_{021}\eta_{022}\eta_{110}+\eta_{021}\eta_{031}\eta_{101}+\eta_{021}\eta_{040}\eta_{110}
\\&+2\eta_{021}\eta_{101}\eta_{211}-\eta_{021}\eta_{110}\eta_{220}-2\eta_{021}\eta_{112}\eta_{200}+\eta_{021}\eta_{130}\eta_{200}-\eta_{022}\eta_{030}\eta_{101}
\\&-\eta_{030}\eta_{031}\eta_{110}+\eta_{030}\eta_{101}\eta_{220}-\eta_{030}\eta_{121}\eta_{200}+\eta_{101}\eta_{102}\eta_{112}-2\eta_{101}\eta_{103}\eta_{111}
\\&+2\eta_{101}\eta_{111}\eta_{301}-2\eta_{101}\eta_{112}\eta_{120}-\eta_{101}\eta_{120}\eta_{130}+2\eta_{101}\eta_{120}\eta_{310}-\eta_{101}\eta_{201}\eta_{211}
\\&-\eta_{101}\eta_{202}\eta_{210}-2\eta_{101}\eta_{210}\eta_{220}+\eta_{101}\eta_{210}\eta_{400}-\eta_{101}\eta_{300}\eta_{310}+\eta_{102}\eta_{103}\eta_{110}
\\&+2\eta_{102}\eta_{110}\eta_{121}-2\eta_{102}\eta_{110}\eta_{301}+2\eta_{102}\eta_{200}\eta_{211}+2\eta_{110}\eta_{111}\eta_{130}-2\eta_{110}\eta_{111}\eta_{310}
\\&-\eta_{110}\eta_{120}\eta_{121}+2\eta_{110}\eta_{201}\eta_{202}+\eta_{110}\eta_{201}\eta_{220}-\eta_{110}\eta_{201}\eta_{400}+\eta_{110}\eta_{210}\eta_{211}
\\&+\eta_{110}\eta_{300}\eta_{301}-2\eta_{111}\eta_{200}\eta_{202}+2\eta_{111}\eta_{200}\eta_{220}-2\eta_{120}\eta_{200}\eta_{211}+\eta_{200}\eta_{201}\eta_{310}
\\&-\eta_{200}\eta_{210}\eta_{301};
\\
\\S_{5}=\ &-2\eta_{011}^2\eta_{012}\eta_{301}+2\eta_{011}^2\eta_{013}\eta_{300}+2\eta_{011}^2\eta_{021}\eta_{310}-2\eta_{011}^2\eta_{031}\eta_{300}+2\eta_{011}^2\eta_{102}\eta_{211}
\\&-2\eta_{011}^2\eta_{103}\eta_{210}+4\eta_{011}^2\eta_{111}\eta_{202}-4\eta_{011}^2\eta_{111}\eta_{220}-4\eta_{011}^2\eta_{112}\eta_{201}-2\eta_{011}^2\eta_{120}\eta_{211}
\\&+4\eta_{011}^2\eta_{121}\eta_{210}+2\eta_{011}^2\eta_{130}\eta_{201}-\eta_{003}\eta_{130}\eta_{200}^2+3\eta_{012}\eta_{121}\eta_{200}^2-3\eta_{021}\eta_{112}\eta_{200}^2
\\&+\eta_{030}\eta_{103}\eta_{200}^2+\eta_{003}\eta_{020}^2\eta_{310}-\eta_{013}\eta_{020}^2\eta_{300}-3\eta_{020}^2\eta_{102}\eta_{211}+3\eta_{020}^2\eta_{112}\eta_{201}
\\&-\eta_{002}^2\eta_{030}\eta_{301}+\eta_{002}^2\eta_{031}\eta_{300}+3\eta_{002}^2\eta_{120}\eta_{211}-3\eta_{002}^2\eta_{121}\eta_{210}-2\eta_{012}\eta_{101}^2\eta_{121}
\\&+2\eta_{013}\eta_{101}^2\eta_{120}+4\eta_{021}\eta_{101}^2\eta_{112}-2\eta_{021}\eta_{101}^2\eta_{310}-4\eta_{022}\eta_{101}^2\eta_{111}-2\eta_{030}\eta_{101}^2\eta_{103}
\\&+2\eta_{030}\eta_{101}^2\eta_{301}+2\eta_{031}\eta_{101}^2\eta_{102}+4\eta_{101}^2\eta_{111}\eta_{220}-4\eta_{101}^2\eta_{120}\eta_{211}+2\eta_{101}^2\eta_{121}\eta_{210}
\\&-2\eta_{101}^2\eta_{130}\eta_{201}-4\eta_{012}\eta_{110}^2\eta_{121}+2\eta_{012}\eta_{110}^2\eta_{301}-2\eta_{013}\eta_{110}^2\eta_{120}+2\eta_{021}\eta_{110}^2\eta_{112}
\\&+4\eta_{022}\eta_{110}^2\eta_{111}-2\eta_{031}\eta_{102}\eta_{110}^2+4\eta_{102}\eta_{110}^2\eta_{211}+2\eta_{103}\eta_{110}^2\eta_{210}-4\eta_{110}^2\eta_{111}\eta_{202}
\\&-2\eta_{110}^2\eta_{112}\eta_{201}+2\eta_{003}\eta_{110}^2\eta_{130}-2\eta_{003}\eta_{110}^2\eta_{310}+\eta_{020}\eta_{101}\eta_{112}\eta_{300}
\\&-2\eta_{020}\eta_{101}\eta_{201}\eta_{211}-\eta_{020}\eta_{101}\eta_{202}\eta_{210}-3\eta_{020}\eta_{102}\eta_{110}\eta_{301}+2\eta_{020}\eta_{102}\eta_{200}\eta_{211}
\\&-\eta_{020}\eta_{103}\eta_{110}\eta_{300}+\eta_{020}\eta_{103}\eta_{200}\eta_{210}+3\eta_{020}\eta_{110}\eta_{201}\eta_{202}-2\eta_{020}\eta_{111}\eta_{200}\eta_{202}
\\&-\eta_{020}\eta_{112}\eta_{200}\eta_{201}-\eta_{200}\eta_{002}\eta_{012}\eta_{121}+\eta_{200}\eta_{002}\eta_{013}\eta_{120}+2\eta_{200}\eta_{002}\eta_{021}\eta_{112}
\\&-2\eta_{200}\eta_{002}\eta_{022}\eta_{111}-\eta_{200}\eta_{002}\eta_{030}\eta_{103}+\eta_{200}\eta_{002}\eta_{031}\eta_{102}+\eta_{200}\eta_{003}\eta_{011}\eta_{121}
\\&+\eta_{200}\eta_{003}\eta_{020}\eta_{130}-\eta_{200}\eta_{003}\eta_{031}\eta_{101}-\eta_{200}\eta_{003}\eta_{040}\eta_{110}+3\eta_{200}\eta_{003}\eta_{110}\eta_{220}
\end{aligned}
\end{equation*}
\begin{equation*}
\begin{aligned}
\\&-\eta_{200}\eta_{004}\eta_{011}\eta_{120}+\eta_{200}\eta_{004}\eta_{030}\eta_{101}-2\eta_{200}\eta_{011}\eta_{012}\eta_{112}-\eta_{200}\eta_{011}\eta_{012}\eta_{130}
\\&+2\eta_{200}\eta_{011}\eta_{013}\eta_{111}+\eta_{200}\eta_{011}\eta_{021}\eta_{103}+2\eta_{200}\eta_{011}\eta_{021}\eta_{121}-\eta_{200}\eta_{011}\eta_{022}\eta_{102}
\\&+\eta_{200}\eta_{011}\eta_{022}\eta_{120}-\eta_{200}\eta_{011}\eta_{030}\eta_{112}-2\eta_{200}\eta_{011}\eta_{031}\eta_{111}+\eta_{200}\eta_{011}\eta_{040}\eta_{102}
\\&-2\eta_{200}\eta_{012}\eta_{020}\eta_{121}+3\eta_{200}\eta_{012}\eta_{022}\eta_{101}+3\eta_{200}\eta_{012}\eta_{031}\eta_{110}-3\eta_{200}\eta_{012}\eta_{101}\eta_{220}
\\&-6\eta_{200}\eta_{012}\eta_{110}\eta_{211}-\eta_{200}\eta_{013}\eta_{020}\eta_{120}-3\eta_{200}\eta_{013}\eta_{021}\eta_{101}+\eta_{200}\eta_{013}\eta_{030}\eta_{110}
\\&+\eta_{200}\eta_{020}\eta_{021}\eta_{112}+2\eta_{200}\eta_{020}\eta_{022}\eta_{111}-\eta_{200}\eta_{020}\eta_{031}\eta_{102}-3\eta_{200}\eta_{021}\eta_{022}\eta_{110}
\\&+6\eta_{200}\eta_{021}\eta_{101}\eta_{211}+3\eta_{200}\eta_{021}\eta_{110}\eta_{202}-3\eta_{200}\eta_{030}\eta_{101}\eta_{202}+3\eta_{200}\eta_{101}\eta_{102}\eta_{130}
\\&-6\eta_{200}\eta_{101}\eta_{111}\eta_{121}+3\eta_{200}\eta_{101}\eta_{112}\eta_{120}-3\eta_{200}\eta_{102}\eta_{110}\eta_{121}-3\eta_{200}\eta_{103}\eta_{110}\eta_{120}
\\&+6\eta_{200}\eta_{110}\eta_{111}\eta_{112}-\eta_{002}\eta_{011}\eta_{030}\eta_{310}+\eta_{002}\eta_{011}\eta_{040}\eta_{300}+3\eta_{002}\eta_{011}\eta_{120}\eta_{220}
\\&-3\eta_{002}\eta_{011}\eta_{130}\eta_{210}+\eta_{002}\eta_{020}\eta_{021}\eta_{310}-\eta_{002}\eta_{020}\eta_{031}\eta_{300}-2\eta_{002}\eta_{020}\eta_{111}\eta_{220}
\\&-\eta_{002}\eta_{020}\eta_{120}\eta_{211}+2\eta_{002}\eta_{020}\eta_{121}\eta_{210}+\eta_{002}\eta_{020}\eta_{130}\eta_{201}-\eta_{002}\eta_{021}\eta_{110}\eta_{220}
\\&+\eta_{002}\eta_{021}\eta_{110}\eta_{400}-\eta_{002}\eta_{021}\eta_{200}\eta_{310}-\eta_{002}\eta_{030}\eta_{101}\eta_{400}+\eta_{002}\eta_{030}\eta_{110}\eta_{211}
\\&+\eta_{002}\eta_{030}\eta_{200}\eta_{301}+\eta_{002}\eta_{031}\eta_{110}\eta_{210}-\eta_{002}\eta_{040}\eta_{110}\eta_{201}+3\eta_{002}\eta_{101}\eta_{120}\eta_{310}
\\&+\eta_{002}\eta_{101}\eta_{130}\eta_{300}-3\eta_{002}\eta_{101}\eta_{210}\eta_{220}+2\eta_{002}\eta_{110}\eta_{111}\eta_{130}-2\eta_{002}\eta_{110}\eta_{111}\eta_{310}
\\&-2\eta_{002}\eta_{110}\eta_{120}\eta_{121}-\eta_{002}\eta_{110}\eta_{120}\eta_{301}-\eta_{002}\eta_{110}\eta_{121}\eta_{300}+\eta_{002}\eta_{110}\eta_{201}\eta_{220}
\\&+2\eta_{002}\eta_{110}\eta_{210}\eta_{211}+2\eta_{002}\eta_{111}\eta_{200}\eta_{220}-2\eta_{002}\eta_{120}\eta_{200}\eta_{211}+\eta_{002}\eta_{121}\eta_{200}\eta_{210}
\\&-\eta_{002}\eta_{130}\eta_{200}\eta_{201}-\eta_{002}\eta_{012}\eta_{020}\eta_{301}+\eta_{002}\eta_{013}\eta_{020}\eta_{300}+\eta_{002}\eta_{020}\eta_{102}\eta_{211}
\\&-\eta_{002}\eta_{020}\eta_{103}\eta_{210}+2\eta_{002}\eta_{020}\eta_{111}\eta_{202}-2\eta_{002}\eta_{020}\eta_{112}\eta_{201}+\eta_{003}\eta_{011}\eta_{020}\eta_{301}
\\&-\eta_{003}\eta_{020}\eta_{101}\eta_{211}+\eta_{003}\eta_{020}\eta_{110}\eta_{400}-\eta_{003}\eta_{020}\eta_{200}\eta_{310}-\eta_{004}\eta_{011}\eta_{020}\eta_{300}
\\&+\eta_{004}\eta_{020}\eta_{101}\eta_{210}-3\eta_{011}\eta_{012}\eta_{020}\eta_{310}+3\eta_{011}\eta_{020}\eta_{022}\eta_{300}-3\eta_{011}\eta_{020}\eta_{102}\eta_{202}
\\&+3\eta_{011}\eta_{020}\eta_{102}\eta_{220}+3\eta_{011}\eta_{020}\eta_{103}\eta_{201}+6\eta_{011}\eta_{020}\eta_{111}\eta_{211}-3\eta_{011}\eta_{020}\eta_{112}\eta_{210}
\\&-6\eta_{011}\eta_{020}\eta_{121}\eta_{201}+\eta_{012}\eta_{020}\eta_{101}\eta_{202}-\eta_{012}\eta_{020}\eta_{101}\eta_{400}+\eta_{012}\eta_{020}\eta_{200}\eta_{301}
\\&-\eta_{013}\eta_{020}\eta_{101}\eta_{201}+2\eta_{020}\eta_{101}\eta_{102}\eta_{112}+\eta_{020}\eta_{101}\eta_{102}\eta_{310}-2\eta_{020}\eta_{101}\eta_{103}\eta_{111}
\\&+2\eta_{020}\eta_{101}\eta_{111}\eta_{301}+2\eta_{011}\eta_{102}\eta_{110}\eta_{310}-2\eta_{011}\eta_{102}\eta_{200}\eta_{220}+4\eta_{011}\eta_{110}\eta_{111}\eta_{301}
\\&+2\eta_{011}\eta_{110}\eta_{112}\eta_{300}-4\eta_{011}\eta_{110}\eta_{201}\eta_{211}-2\eta_{011}\eta_{110}\eta_{202}\eta_{210}-2\eta_{011}\eta_{112}\eta_{200}\eta_{210}
\\&+2\eta_{011}\eta_{120}\eta_{200}\eta_{202}+2\eta_{011}\eta_{121}\eta_{200}\eta_{201}+3\eta_{002}\eta_{011}\eta_{021}\eta_{301}-3\eta_{002}\eta_{011}\eta_{022}\eta_{300}
\\&-6\eta_{002}\eta_{011}\eta_{111}\eta_{211}+6\eta_{002}\eta_{011}\eta_{112}\eta_{210}-3\eta_{002}\eta_{011}\eta_{120}\eta_{202}+3\eta_{002}\eta_{011}\eta_{121}\eta_{201}
\\&-2\eta_{011}\eta_{012}\eta_{110}\eta_{400}+2\eta_{011}\eta_{012}\eta_{200}\eta_{310}+2\eta_{011}\eta_{021}\eta_{101}\eta_{400}-2\eta_{011}\eta_{021}\eta_{200}\eta_{301}
\\&-4\eta_{011}\eta_{101}\eta_{111}\eta_{310}-2\eta_{011}\eta_{101}\eta_{120}\eta_{301}-2\eta_{011}\eta_{101}\eta_{121}\eta_{300}+2\eta_{011}\eta_{101}\eta_{201}\eta_{220}
\\&+4\eta_{011}\eta_{101}\eta_{210}\eta_{211}+4\eta_{110}\eta_{011}\eta_{031}\eta_{201}-3\eta_{110}\eta_{020}\eta_{022}\eta_{201}+4\eta_{110}\eta_{101}\eta_{121}\eta_{201}
\\&+3\eta_{110}\eta_{012}\eta_{020}\eta_{211}+3\eta_{110}\eta_{013}\eta_{020}\eta_{210}-6\eta_{110}\eta_{020}\eta_{111}\eta_{112}-4\eta_{110}\eta_{011}\eta_{021}\eta_{211}
\\&-4\eta_{110}\eta_{021}\eta_{101}\eta_{301}+2\eta_{110}\eta_{002}\eta_{012}\eta_{211}-2\eta_{110}\eta_{002}\eta_{021}\eta_{202}+2\eta_{110}\eta_{002}\eta_{022}\eta_{201}
\\&-2\eta_{110}\eta_{002}\eta_{102}\eta_{121}+2\eta_{110}\eta_{011}\eta_{012}\eta_{202}+4\eta_{110}\eta_{012}\eta_{101}\eta_{310}+2\eta_{110}\eta_{003}\eta_{101}\eta_{121}
\\&-4\eta_{110}\eta_{012}\eta_{101}\eta_{112}+2\eta_{110}\eta_{021}\eta_{101}\eta_{103}-4\eta_{110}\eta_{101}\eta_{102}\eta_{220}+4\eta_{110}\eta_{101}\eta_{120}\eta_{202}
\\&-4\eta_{110}\eta_{011}\eta_{103}\eta_{111}+4\eta_{110}\eta_{013}\eta_{101}\eta_{111}-4\eta_{110}\eta_{011}\eta_{102}\eta_{130}+6\eta_{110}\eta_{020}\eta_{102}\eta_{121}
\\&-2\eta_{110}\eta_{022}\eta_{101}\eta_{102}+2\eta_{110}\eta_{002}\eta_{103}\eta_{120}-2\eta_{110}\eta_{004}\eta_{101}\eta_{120}+4\eta_{110}\eta_{011}\eta_{112}\eta_{120}
\\&-3\eta_{110}\eta_{003}\eta_{020}\eta_{220}-4\eta_{110}\eta_{011}\eta_{022}\eta_{210}+4\eta_{110}\eta_{011}\eta_{102}\eta_{112}-2\eta_{110}\eta_{003}\eta_{011}\eta_{211}
\\&+4\eta_{110}\eta_{011}\eta_{012}\eta_{220}-2\eta_{110}\eta_{011}\eta_{013}\eta_{201}-2\eta_{110}\eta_{002}\eta_{013}\eta_{210}+2\eta_{110}\eta_{004}\eta_{011}\eta_{210}
\\&-4\eta_{110}\eta_{101}\eta_{112}\eta_{210}-2\eta_{101}\eta_{012}\eta_{110}\eta_{130}+4\eta_{101}\eta_{021}\eta_{110}\eta_{121}-2\eta_{101}\eta_{030}\eta_{110}\eta_{112}
\end{aligned}
\end{equation*}
\begin{equation*}
\begin{aligned}
\\&+2\eta_{101}\eta_{011}\eta_{030}\eta_{211}+2\eta_{101}\eta_{011}\eta_{031}\eta_{210}-4\eta_{101}\eta_{011}\eta_{120}\eta_{121}-4\eta_{101}\eta_{011}\eta_{021}\eta_{202}
\\&+4\eta_{101}\eta_{011}\eta_{022}\eta_{201}+3\eta_{101}\eta_{002}\eta_{022}\eta_{210}-4\eta_{101}\eta_{011}\eta_{013}\eta_{210}-2\eta_{101}\eta_{011}\eta_{040}\eta_{201}
\\&+2\eta_{101}\eta_{020}\eta_{031}\eta_{201}+2\eta_{101}\eta_{012}\eta_{020}\eta_{220}-2\eta_{101}\eta_{020}\eta_{021}\eta_{211}-2\eta_{101}\eta_{020}\eta_{022}\eta_{210}
\\&+2\eta_{101}\eta_{020}\eta_{112}\eta_{120}-2\eta_{101}\eta_{011}\eta_{021}\eta_{220}+3\eta_{101}\eta_{002}\eta_{030}\eta_{202}-3\eta_{101}\eta_{002}\eta_{021}\eta_{211}
\\&-3\eta_{101}\eta_{002}\eta_{031}\eta_{201}+6\eta_{101}\eta_{002}\eta_{111}\eta_{121}+4\eta_{101}\eta_{011}\eta_{012}\eta_{211}+4\eta_{101}\eta_{011}\eta_{111}\eta_{130}
\\&-4\eta_{101}\eta_{031}\eta_{110}\eta_{111}-4\eta_{101}\eta_{011}\eta_{102}\eta_{121}-2\eta_{101}\eta_{020}\eta_{102}\eta_{130}+2\eta_{101}\eta_{040}\eta_{102}\eta_{110}
\\&-6\eta_{101}\eta_{002}\eta_{112}\eta_{120}+4\eta_{101}\eta_{011}\eta_{103}\eta_{120}+2\eta_{101}\eta_{022}\eta_{110}\eta_{120}.
\end{aligned}
\end{equation*}

\end{document}